# Venus as an Analog for Hot Earths

Giada Arney[1,2,3,*], Stephen Kane[4]

[1]Planetary Systems Laboratory, NASA Goddard Space Flight Center, Greenbelt, MD
[2]Virtual Planetary Laboratory, University of Washington, Seattle, WA
[3]Sellers Exoplanet Environments Collaboration, NASA Goddard Space Flight Center, Greenbelt, MD
[4]Department of Earth Sciences, University of California at Riverside, Riverside, CA

* Corresponding author: giada.n.arney@nasa.gov

**Abstract**
Here, we evaluate our nearest planetary neighbor, Venus, as an exemplar of the runaway greenhouse state that bounds the inner edge of the habitable zone. Despite its current hellish surface environment, Venus may once have been habitable with oceans of surface liquid water. Over time, it lost its potentially clement environment as the sun brightened. Today, it represents the end-state of habitable planet evolution, and it therefore provides valuable lessons on habitability as a planetary process. Beyond the solar system, exo-Venus analogs are likely common types of planets, and we likely have already discovered many of Venus' sisters orbiting other stars. Furthermore, our near-future exoplanet detection and characterization methods are biased towards observing them. Therefore, it is instructive to consider what Venus can teach us about exo-Venus analogs. By observing exo-Venus planets of differing ages in in different astrophysical contexts, these distant hot terrestrial worlds may likewise allow us to witness processes that occurred on Venus in the past.

## 1 Introduction

Although Venus is currently one of the most uninhabitable places in the solar system, understanding our nearest planetary neighbor may unveil general processes of how planetary environments evolve – and can lose habitability – over time. In many ways, Venus is the most Earth-like planet in the solar system in terms of its size, mass, and bulk composition, yet the surface conditions of these two worlds could not be more different. Venus has the hottest terrestrial surface in the solar system. Despite this, Venus may once have hosted clement conditions in its deep past, possibly even with a liquid surface water ocean. However, only fragments remain from this hypothetical lost paradise that researchers have pieced together into a tantalizing yet incomplete picture.

Better understanding the evolution of the Venusian environment may inform our understanding of similar exoplanets. However, understanding the processes that affect habitability will be particularly challenging in the context of characterizing these distant and data-limited worlds around other stars. The Venusian environment is a natural laboratory next door to study the complex processes that shape habitability on hot terrestrial worlds, providing empirical data to help us better understand the thin line that separates habitable worlds from uninhabitable ones. Photochemical processes that occur on Venus may also help us to understand the potential for mechanisms to generate abiotic oxygen in exoplanet atmospheres, which has important implications for our understanding of oxygen as a robust biosignature.

In addition to informing our understanding of planetary habitability, ground-truth information from "our" Venus can also inform and constrain our interpretations of data from exo-Venus analogs. Important for future observations of exoplanets, Venus-analogs represent one of the most readily observable types of terrestrial planets for the transit transmission observations that will become possible in the near future with the James Webb Space Telescope. This is because Venus-analog worlds orbit closely to their stars and therefore enjoy a higher transit probability and frequency of transit events compared to planets in the habitable zone. Indeed, many terrestrial exoplanets already discovered, and undoubtedly many that will be uncovered by TESS, are likely to be more Venus-like than Earth-like.

Here, we will tell the story of Venus, starting with a description her present diabolical state, continuing to her possible history and evolution that may include past habitable conditions, and ending with considerations of "pale yellow dots" whirling around distant stars.

## 2 The current state of Venus

Venus is a complex and intriguing planet, and its surface conditions were veiled in mystery for many decades, prompting wild speculations of swampy oases beneath its dense butter-colored clouds (e.g. Arrhenius 1918). A hint of the hot Venus surface temperature came from millimeter wave observations in the late 1950s (Mayer et al. 1958), but confirmation of the true hellish nature of Earth's twisted sister had to wait for spacecraft visits in the 1960s (e.g. Dickel 1966). We now know that the surface temperature of Venus is hot enough to melt lead, so it is of little surprise that the few landers that have been to the Venusian surface have survived for no more than a few hours at most. Yet the challenges of studying the Venus environment are also opportunities that invite us to consider how and why this planet that is in many ways similar to Earth in terms of size (0.94 x Earth's radius), mass (0.81 x Earth's mass), and bulk composition ended up so different. These questions are even more important in light of evidence suggesting that Venus was once more Earth-like. How did Venus lose its habitability? When? Why?

Today, Venus is a planet of many extremes. At its surface, a crushing 92 bars of pressure bear down, equivalent to the pressure at 1.6 km underwater on Earth. The clouds are composed not of water but of a concentrated solution of sulfuric acid ($H_2SO_4$) and water. Its globally averaged surface temperature is 735 K (Marov et al. 1973; Seiff et al. 1985), although higher and lower elevations are colder and hotter, respectively. At its lowest point in Diana Chasma, the surface is a scorching 765 K, so hot that a faint red incandescence emanates from the ground (Bullock and Grinspoon 2013). Even Venus' rotation is unusual in the solar system. Venus has a rotation period of 116.75 days, and it orbits sun every 225 days, making its sidereal day 243 days, longer than its year. Possibly, atmospheric tides in the Venusian atmosphere led to present rotation rate (Dobrovolskis and Ingersoll 1980), which is retrograde relative to the other planets. It is unclear why Venus rotates in this manner, which is unlike every other solar system planet. Correia & Laskar (2003) suggest that its rotational axis could have rotated 180° due to internal core-mantle friction and tides from the atmosphere. They suggest that for planets with thick atmospheres like Venus, most initial conditions can lead to Venus-like rotation regimes via two mechanisms: the spin axis can flip directions, or a planet with prograde rotation can become retrograde as the obliquity decreases towards zero. Alternatively, Venus' unusual rotation may be due to an impact by a large planetesimal (Baines et al 2013). This slow rotation may be responsible for Venus' lack of internally-generated magnetic field (Stevenson 2003).

The most obvious characteristic of Venus is its temperature. The solar constant (i.e. how much solar radiation a planet receives per unit area) at Venus is almost two times as much Earth's solar constant (2610 W/m$^2$ compared to Earth's 1360 W/m$^2$) but this is not the only cause of its hot surface temperatures. Its high albedo due to its brightly reflective $H_2SO_4$ clouds actually mean that less solar radiation is absorbed by the Venus atmosphere than Earth absorbs: Earth reflects about a third of the incoming radiation back to space, while Venus reflects over 70%. Incredibly, only about 2.5% of the total incoming solar radiation actually reaches the surface, as the rest is absorbed by the atmosphere. The high surface temperatures Venus experiences are due to its incredibly powerful greenhouse effect.

## 2.1 The Atmosphere

The Venus atmosphere is extremely efficient at trapping heat: the dense 92 bar atmosphere is composed of 96.5% $CO_2$, producing an almost impenetrable greenhouse enabled by strong, pressure-broadened, infrared absorption bands. Carbon dioxide produces its strongest absorption bands near 4.4 and 15 µm. On Earth, the 15 µm $CO_2$ band is important for our greenhouse effect. Venus' elevated temperature means that it still emits significant amounts of radiation at wavelengths as short as 4.4 µm, and so this band is more important for the Venusian greenhouse than it is for Earth's. Collisional broadening of these $CO_2$ bands increases their extent in wavelength, and broadened $H_2O$ bands from trace water vapor in the atmosphere are effective at closing gaps between $CO_2$ bands where radiation could otherwise escape. It is only through narrow spectral windows in the near-infrared between roughly 1-2.5 µm that thermal radiation can leak out to space, mostly through a spectral window near 2.3 µm. About 3% of outgoing thermal radiation is able to escape between 2.3 and 2.7 µm (Pollack 1969; Pollack et al. 1980). Figure 1 shows the vertical profiles of key trace gases from the Venus International Reference Atmosphere (Moroz & Zasova 1997).

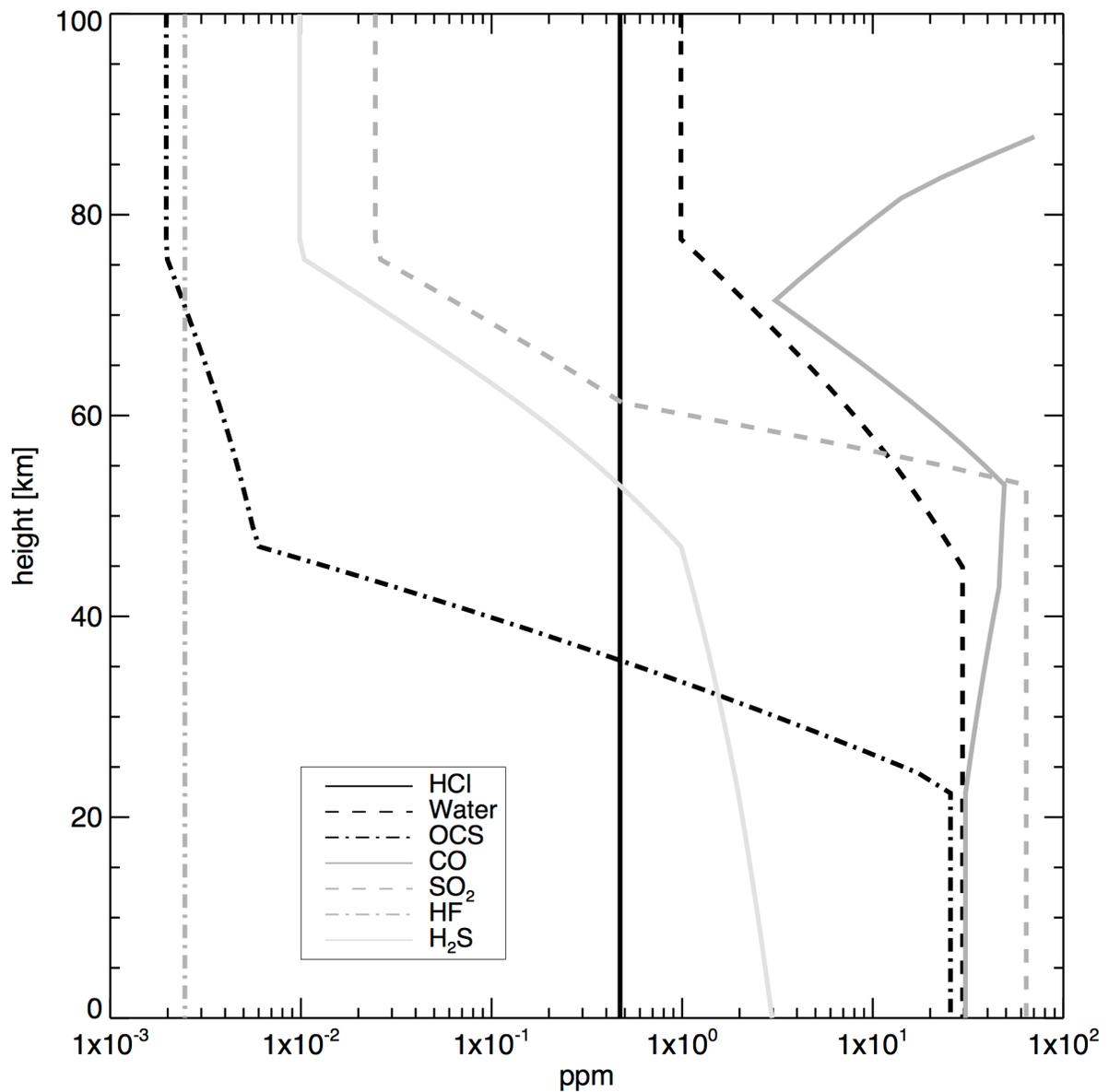

**Figure 1.** Profiles of trace gases in the Venus atmosphere. $CO_2$, not shown in this figure, comprises about 96.5% of the Venus atmosphere, and $N_2$ (also not shown) comprises 3.5%.

Molecular nitrogen is the second most abundant gas in the Venus atmosphere, comprising 3.5% of the total. Although this is a small fraction of the whole Venusian atmosphere, it is about 4 times as much total $N_2$ as in the Earth's atmosphere (Hoffman et al. 1980; Von Zahn et al. 1983). On Earth, nitrogen cycles between the atmosphere and mantle driven by various biotic and abiotic processes (e.g. Canfield et al. 2010; Wordsworth 2016). On Venus, it has been proposed that oxygen liberated by $H_2O$ photolysis and water loss (described in Section 3) could have been absorbed by the mantle, increasing its oxygen fugacity and releasing $N_2$ to the atmosphere (Wordsworth and Pierrehumbert 2014). On Earth, most of its $CO_2$ is locked up on carbonates on continental margins. If Earth's crustal $CO_2$ and $N_2$ was liberated into the atmosphere, our planet would have about 100 bars of $CO_2$ and 0.7 bars of $N_2$ (Sleep & Zahnle 2001; Zahnle et al 2007), similar to Venus. Thus, Venus appears to have degassed its bulk inventory of $CO_2$ and $N_2$ to the atmosphere.

The atmosphere of Venus super-rotates, meaning that upper layers rotate faster than the surface. Extreme winds travel from east to west at speeds 60 times faster than the surface rotation rate (Schubert 1983). These speeds reach 120 m/s near the equatorial cloud tops. The complete mechanism that causes the Venus super-rotation is not well understood, but evidence suggests angular momentum is transferred between the planetary surface and upper atmosphere (Mendonça & Read 2016). Studies of atmospheric waves propagating in the Venusian atmosphere may be key to understanding the nature of its super-rotation, and such waves have been observed in the Venus atmosphere (Peralta et al. 2009, 2014; Piccialli et al. 2014).

Studies of these waves have been enabled by the JAXA Akatsuki mission, which is currently in orbit around Venus. Akatsuki was intended to go into orbit around Venus in 2010, but the insertion attempted failed. Instead, it remained in a heliocentric orbit until 2015, when a second close to pass to Venus resulted in a successful orbital insertion. Akatsuki data has revealed an unusual stationary wave in the upper atmosphere (Fukuhara et al. 2017). Normal wind speeds at these altitudes whip across the planet at roughly 100 m/s, but this UV-bright, bow-shaped feature remains stationary relative to the surface far below. The center of the feature is located above the western slope of highland region Aphrodite Terra and may be a stationary atmospheric gravity wave associated with lower atmosphere wind flows over this feature. On Earth, gravity waves in the troposphere can be generated when wind flows over mountainous regions; waves propagating through the atmosphere at higher altitudes where the air is thinner (i.e. over mountains) can experience nonlinear wave breaking effects, affecting the dynamics of Earth's middle atmosphere. On Venus, however, the propagation of such mountain waves to the cloud tops may be difficult as convection between the cloud top and ground can disturb wave features (e.g. Seiff et al. 1980). If the feature seen by Akatsuki is indeed associated with Aphrodite Terra, this implies the atmospheric dynamics of the deep Venusian atmosphere may be more complex and less well understood than previously thought.

The axial tilt of Venus is small: its obliquity is only 1.7 degrees, compared to Earth's 23.5 degrees. Thus, any seasonal variability in the atmosphere should be extremely minor. Indeed, the temperature gradient between the equator and poles is only ~4 K, with little annual variation (Seiff et al. 1980). Despite this apparent lack of seasonality, surprising variability has been observed in distributions of gases the Venus atmosphere, which vary on unknown timescales due to unknown mechanisms. Hemispherical dichotomies in the abundances of trace gases like $H_2O$, $CO$, $SO_2$, and $OCS$ have been observed to vary on poorly constrained timescales (Arney et al. 2014; Collard et al. 1993; Krasnopolsky 2010a; Marcq et al. 2006). For example, Arney et al (2014) found time-varying hemispherical dichotomies in the distributions of $CO$, $H_2O$, $OCS$, $SO_2$, and cloud droplet $H_2SO_4$ concentration using ground-based observations of the Venus sub-cloud atmosphere in observations from 2009 and 2010. It is known that the distribution of CO in the lower atmosphere varies with latitude due to atmospheric circulation processes. CO is thought to be produced from $CO_2$ photolysis above the cloud deck, and it freshly downwells near 60 degrees north and south. Atmospheric circulation carries it to the equator at lower elevations, and reactions with surface minerals or other processes can transform it into OCS. Indeed, observations suggest that CO and OCS are anticorrelated to each other, with CO abundances peaking at higher latitudes and reaching minima at lower latitudes, while OCS displays the opposite behavior (Arney et al. 2014; Marcq et al. 2005, 2006, 2008), and so their observed opposing hemispherical dichotomies may be chemically related to these processes. Seasonal hemispherical dichotomies have been observed on other solar system bodies such as Saturn (Sinclair et al. 2013) and Titan (e.g. Coustenis & Bezard 1995; Teanby et al. 2008). However, as mentioned, the small Venusian obliquity makes seasonally-driven variations difficult to invoke as an explanation for the observed variable dichotomies.

Other variations in the Venus atmosphere have been observed besides variable hemispherical dichotomies. For example, diurnal variations may be present (Sandor et al. 2010), and cloud top wind speeds were observed to increase between 2006 and 2012 in Venus Express observations (Khatuntsev et al. 2013), and periodicity in wind speed oscillations tracked by cloud motion has been reported (Kouyama et al. 2013).

## 2.2 The clouds
The Venus cloud deck obscures the surface at most wavelengths and is composed primarily of $H_2SO_4$ droplets that form photochemically in the upper layers at altitudes around 60 km (160 mbar) through reactions involving $SO_2$ and trace amounts of water vapor (Yung and DeMore 1982):

$$SO_2 + HO_2 \rightarrow SO_2 + OH$$
$$SO_3 + H_2O + M \rightarrow H_2SO_4 + M$$

At these high altitudes where it forms, $H_2SO_4$ exists as a fine haze. By diffusion, the $H_2SO_4$ can then condense into the main cloud deck below. Cloud droplets are composed of a concentrated $H_2SO_4/H_2O$ solution and are typically 75-85% $H_2SO_4$ (Barstow et al. 2012; Cottini et al. 2012). The main cloud deck can be divided into three major layers characterized by different sized particles, plus a fine haze above and below the main clouds (Crisp 1986; Table 1). The upper layers (>70 km) are composed of a fine haze of "mode 1" particles approximately 0.4 µm in radius. Below this, the upper cloud (57-70 km) contains "mode 2" particles, approximately 1.4 µm in radius; mode 1 particles are also in the upper cloud. In the middle (50-57 km) and lower clouds (48-50 km), the cloud particles consist of mode 2 and mode 3 particles, which have an average radius of 3.85 µm (Knollenberg and Hunten 1980). These modes are summarized in Table 2. The large mode 3 particles are responsible for the bulk of the opacity and mass of the cloud layers (Crisp 1986). Mode 3 particles may have a crystalline component (Esposito et al. 1983; Knollenberg and Hunten 1980), requiring the presence of some species other than $H_2SO_4$. It has also been proposed that mode 3 particles represent the large end of the particle size distribution for the mode 2 particles (Toon et al. 1984). In this case, the particles would be non-crystalline, but their optical properties would disagree with Pioneer Venus nephelometer (Ragent and Blamont 1980) and solar flux radiometer data (Tomasko et al. 1980). Models of the cloud structure on Venus have been proposed by several authors (Crisp 1986; Grinspoon et al. 1993; J. Pollack et al. 1993; Tomasko, Doose, and Smith 1985) and more recent studies have worked to better constrain the cloud structure and properties (e.g. Barstow et al. 2012; Satoh et al. 2009)

Table 1. Venus cloud layer properties. Derived from Esposito et al. (1983).

| Layer name | Altitude (km) | Temperature (K) | Pressure (bar) | Particle mode types (see Table 2) | Optical depth at 0.63 um |
| --- | --- | --- | --- | --- | --- |
| Upper haze | 70-90 | 255-190 | 0.0267-0.0028 | 1 | 0.2-1 |
| Upper cloud | 56.5-70 | 286-255 | 0.406-0.0267 | 1, 2 | 6-8 |
| Middle cloud | 50.5-56.5 | 245-286 | 0.981-0.406 | 1, 2', 3 | 8-10 |
| Lower cloud | 47-5-50.5 | 367-345 | 1.391-0.981 | 1, 2, 3 | 6-12 |
| Lower haze | 38-47.5 | 430-367 | 4-1.39 | 1 | 0.1-0.2 |

Table 2. Venus cloud particle sizes from Crisp (1986).

| Mode | Effective Radius (µm) | Variance |
| --- | --- | --- |
| 1 | 0.49 | 0.22 |
| 2 | 1.04 | 0.19 |
| 2' | 1.4 | 0.207 |
| 3 | 3.85 | 0.262 |

The altitude at which the $H_2SO_4$ vapor pressure exceeds its saturation vapor pressure sets the altitude of the bottom of the cloud deck at about 47 km (about 2 bars). A fine haze of $H_2SO_4$ extends below the cloud deck. Below the clouds $H_2SO_4$ can exist in vapor phase down to about 38 km (about 4 bars), below which it is thermochemically decomposed at temperatures around 430 K (Bullock & Grinspoon 2001) through the reaction $H_2SO_4 \rightarrow SO_3 + H_2O$. The clouds themselves are important to the Venus greenhouse effect. Climate modeling shows that if the cloud deck was removed on Venus, the surface temperature would decrease by 142.8 K

(compare to a decrease of 422.7 K if $CO_2$ is removed, and 68.8 K if $H_2O$ is removed; Bullock & Grinspoon 2001).

Also in the clouds is an "unknown UV absorber" (e.g. Esposito, 1980; Markiewicz et al. 2014) which absorbs wavelengths 320-500 nm and is responsible for the characteristic yellowish color of Venus. The absorber appears to be distributed in the upper cloud layer at 58 to 62 km (Ekonomov et al., 1984). A number of candidates have been suggested for the identity of this absorber including elemental sulfur, $Cl_2$, $SCl_2$, and $S_2O$ (Esposito et al., 1983, 1997), although these compounds do not have absorption spectra that match observations. Additionally, proposed abundances of $Cl_2$, $SCl_2$, and $S_2O$ exceed those predicted by photochemical models by two orders of magnitude. Zasova et al. (1981) proposed that the absorber is a solution of ~1% $FeCl_3$ in $H_2SO_4$, which reasonably matches the observed spectrum of Venus. These iron-bearing particles may be lofted from surface rocks and might act as condensation nuclei for cloud particles in the upper atmosphere (Krasnopolsky, 1989).

Gaseous $SO_2$ can be lost from the atmosphere through reactions with surface minerals, which has implications for the lifetime of the $H_2SO_4$ cloud deck. Fegley and Prinn (1989) showed that at Venus-like conditions, $SO_2$ should react with calcite on the surface to form anhydrite via:
$$CaCO_3 + SO_2 \leftrightarrow CaSO_4 + CO$$
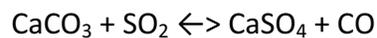

Based on this reaction, all of the $SO_2$ in the Venus atmosphere could be lost in only two million years (Fegley & Prinn 1989). However, this calculation did not account for diffusion of $SO_2$ through surface rock layers to reach fresh calcite, which would increase the atmospheric lifetime of $SO_2$. Bullock and Grinspoon (2001) applied a model that included diffusion to the problem and estimated the atmospheric lifetime of $SO_2$ to be ten times longer than the Fegley and Prinn (1989) calculation: they estimate the lifetime for $SO_2$ to be 20 million years. Bullock and Grinspoon (2001) estimated that amount of $SO_2$ currently in the atmosphere is in excess of equilibrated conditions by a factor of 100. If $SO_2$ concentrations drop below 10 ppmv, the clouds should dissipate (Bullock & Grinspoon 2001). Thus, the existence of a long-lived cloud deck implies ongoing volcanic processes generating $SO_2$, and the existence of the clouds on modern Venus require $SO_2$ injections to the atmosphere within the past 20 million years. The most abundant compounds outgassed by typical basaltic volcanism are $CO_2$, $H_2O$, and $SO_2$ (Kaula & Phillips 1981). Fegley and Prinn (1989) estimate that between 0.4-11 $km_2$ of fresh eruptions are needed to maintain the cloud deck.

Strikingly, variations in $SO_2$ at the cloud top level has been observed across multiple decades of observations by Pioneer Venus (in orbit from 1978 to 1992) and Venus Express (VEx; its tenure at Venus started in 2006 and ended in 2015 when it was de-orbited in the upper atmosphere), suggesting that there may be long-term cycling mechanisms transporting $SO_2$ from the lower atmosphere to the upper layers. This mechanism could be related to atmospheric dynamical and transport processes. The sub-cloud atmosphere contains roughly 130-180 ppm $SO_2$ (Arney et al 2014; Bezard et al. 1993; Pollack et al. 1993; Marcq et al. 2008), while the atmosphere above the clouds contains only a few hundred ppb (Marcq et al 2013; Krasnopolsky 2010b), so transport of the relatively $SO_2$-rich lower layers to higher levels have the potential to cause measurable changes. The variations in $SO_2$ seen above the clouds could also be related to volcanic injections of $SO_2$, rather than purely atmospheric dynamical processes, which is particularly intriguing in light of the necessity of volcanism to maintain the cloud deck, and other evidence for volcanism described below.

## 2.3 The surface and possible volcanism

Little was known about the surface and possible volcanic processes of Venus prior to the Magellan mission, which launched in 1989. Earlier observations using, e.g., the 85-ft Goldstone radio dish in the early 1960s (Carpenter, 1964, 1966; Muhleman 1961) could discern the Venusian solid body rotation rate and its radius. Magellan's radar images mapped nearly the whole surface at a resolution of 200 m, unveiling a basaltic landscape. Most of the Venus surface is rolling plains splattered with a variety of volcanic features and lava flows but without evidence for plate tectonic activity. For instance, rift valleys exist on Venus (e.g. in the West Eistla Region), but these features are not believed to form from plate spreading as they do on Earth. Instead, they probably formed from volcanic activity that stretched and deformed parts of the crust.

Remote searches for active volcanism on the surface of Venus (and remote measurements of the lower atmosphere) are enabled by "near-infrared spectral windows" (Allan & Crawford 1984). These windows were discovered when Allen and Crawford (1984) found an unexpected excess of radiation near 1.74 and 2.3 µm on the Venus nightside, and subsequent studies showed that they sense thermal emission emanating from below the cloud layer (Allen 1987). Subsequently, additional windows at shorter wavelengths were discovered at 1.0, 1.1, 1.18, 1.27, and 1.31 µm (Carlson et al 1991; Crisp et al 1991) that sense the surface and lowest atmospheric scale height (~16 km). In this wavelength region, the clouds are very weakly absorbing (Figure 2), although they are still highly scattering. The windows exist between strong, pressure-broadened $CO_2$ and $H_2O$ bands, allowing thermal radiation to leak to space (Section 2.1). When observed on the planet's nightside, because there is no confounding scattered sunlight, nightside thermal emission from the sub-cloud atmosphere and surface can be sampled directly even from Earth. However, the scattering footprint of radiation escaping through the cloud deck is still about 100 km$^2$ (Drossart et al 2007), so smaller-scale features are smeared out.

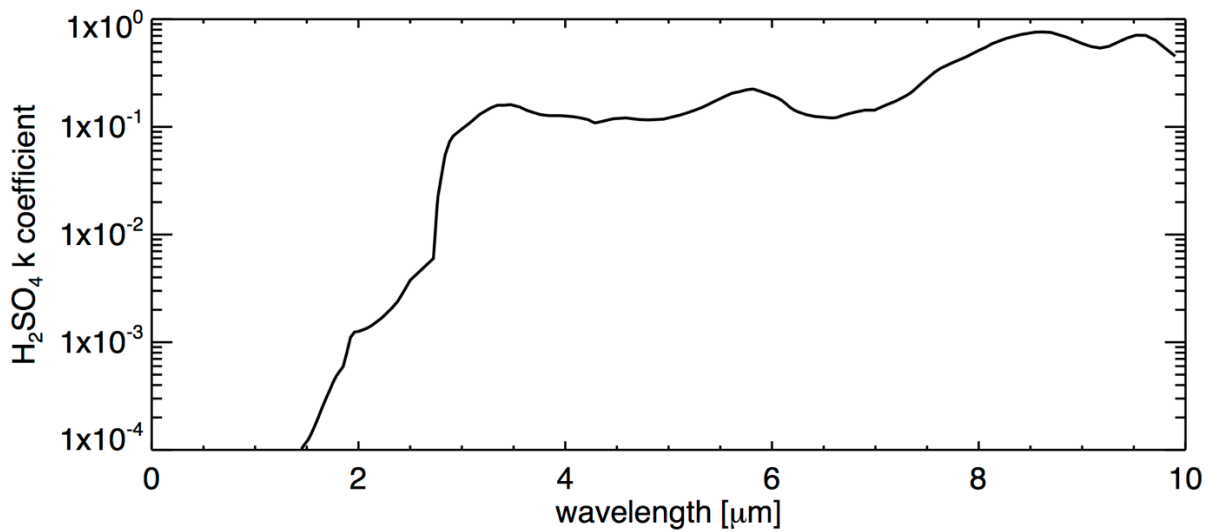

**Figure 2.** The extinction coefficient (k) of a 75% $H_2SO_4$ solution as a function of wavelength from Palmer and Williams (1975). The extinction coefficient decreases by orders of magnitude for wavelengths shorter than about 2.5 µm, allowing thermal radiation to escape at these wavelengths.

Initial searches for volcanism on Venus did not prove fruitful: an analysis of VEx data obtained with the Venus Monitoring Camera (VMC) between October 31, 2007 through June 15, 2009 did not reveal any hot spots that might be indicative of volcanism (Shalygin et al. 2012). However, more recently, Smrekar et al (2010) discussed the discovery of nine emissivity anomalies identified in VEx Visible and Infrared Thermal Imaging Spectrometer (VIRTIS) data as sites of possible volcanic activity which may be 2.5 million years old, and more likely as young as 250,000 years old. These age estimates are derived from estimated weathering rates of fresh basalt. Strengthening the case for a volcanic interpretation, Magellan gravity data suggests the emissivity anomalies are associated with surface regions likely to have a thin, elastic lithosphere. More evidence for modern volcanism came in 2015 via new analyses of VMC observations. Shalygin et al. (2015) observed four temporally variable hotspots at volcanoes Ozza Mons and Maat Mons in the Ganiki Chasma rift zone, implying fresh volcanic activity.

Interestingly, the surface of Venus has only about 1000 randomly distributed impact craters largely unaffected by tectonism or volcanism (Schaber et al. 1998). This cratering record suggests that almost the whole surface of Venus is roughly 300-800 million years old (Strom et al 1994; Phillips et al 1992; Namiki & Solomon 1994; Herrick and Rumpf 2011). This unusual finding has been interpreted to suggest a global resurfacing event (Strom et al 1994) that obliterated older crust from the first 80% of Venus' history. Such an event could be driven by rising mantle temperatures caused by heat trapped by a thick lithosphere that cannot be released because Venus has no plate tectonics. The rising temperatures weaken the crust until widespread subduction occurs and the crust is completely recycled (Schaber et al 1992; Strom et al 2994; Turcotte et al 1999). This resurfacing event could have caused a 100 K change in temperature at the surface (Bullock & Grinspoon 2001), which would have put stresses on the lithosphere over the next 100 million years, possibly creating wrinkle ridge features that have been observed (Solomon et al. 1999).

Intriguingly, studies have suggested that certain regions on the surface called tesserae, which are highly deformed elevated surface regions (e.g. Aphrodite Terra), did not participate in the global resurfacing event (Romeo and Turcotte 2008; Ivanov and Head 1996; 2013). The word "tessera" means "floor tiles" in Greek (such as to construct a mosaic), and the textures of these regions are unique to Venus in the Solar System. These tesserae could record a distinct tectonic regime from the Venusian past (Gilmore et al. 1997; Brown & Grimm 1997), and models suggest that they may have formed by mantle up/downwelling processes (Ghent et al. 2005), or they may represent deformed lava planes (Hansen et al. 2000). If the tessera represent ancient crustal materials, they might someday be sampled to yield clues about the pre-resurfacing history of Venus. Venus Express data hints that tesserae may have a lower emissivity than the low-lying basaltic planes (Basilevsky et al 2012). This could be indicative of composition differences indicative of silica-rich minerals (Hashimoto & Sugita 2003) possibly formed through ancient continent building processes, or alternatively the emissivity variations could indicate, e.g., grain size differences in tesserae (Basilevsky et al 2004; 2007).

Beneath the surface, there is evidence that Venus may still have a liquid core from its tidal Love number from Magellan and Pioneer Venus data (Konopliv and Yoder 1996; Yoder 1997). Despite this, the planet lacks a magnetic field (Russell 1980). If Venus had a magnetic field in the past, it may have been lost if the planet transitioned from a tectonic regime to a sluggish mantle or stagnant lid regime that is inefficient at allowing internal heat to escape. Alternatively, it is thought that the core dynamo on Earth is partially driven by compositional convection caused by the inner core solidifying. If the Venusian core has not cooled enough to allow this, that may stymie the generation of a dynamo (Stevenson et al. 1983).

## 2.4 Suggestions of extant life

Although numerous questions remain about the nature of current Venus, it is abundantly clear that Earth's closest planetary neighbor (closest both in distance and in terms of bulk properties) appears to be one of the least likely terrestrial worlds of the solar system to support life. However, it has been suggested that there may be possible habitats for microbes in the cloud layers where temperatures and pressures are more amenable to life (Schulze-Makuch et al 2002; 2004). For instance, at altitudes of around 50 km, pressures are 1 bar and temperatures are 300-350 K. It is hypothesized Venusian microbes could have taken refuge in this relatively clement atmospheric layer after the surface of Venus was rendered uninhabitable. On Earth, microbes can exist in the upper atmosphere (Smith 2013). It has been hypothesized that the large "mode 3" particles in the Venus clouds might even be biological (Grinspoon 1997), but this possibility remains highly speculative. In any case, it is clear that Venus continues to hold fast to its secrets and further study is needed to unlock its mysteries.

## 3 The History and Evolution of Venus

The hellish current conditions of present day Venus stand in stark contrast to possible warm, watery conditions that have been suggested for its deep past. If it was indeed habitable, how was this paradise lost? Moreover, if Venus had surface liquid water, how much was there, and how long did it last? These questions do not have definitive answers, yet there have been numerous studies investigating aspects of these problems, and they have sketched the outlines of a doomed once-habitable world just next door.  Understanding Venus is vital to understanding processes that shape habitability in our solar system and elsewhere in the universe.

Despite their vast differences today, Venus may have formed with similar bulk composition and initial volatile inventory to Earth. For instance, Morbidelli et al (2012) suggests that Venus and Earth are unlikely to have formed with greater than five orders of magnitude difference in their water inventories, and some accretional modeling scenarios suggest Venus received 5 - 30 Earth oceans worth of water from accretion of volatile-rich bodies (Chassefiere et al 2012). However, in some formation scenarios, most of the water may have been lost in the first 100 Myr of the planet's history (Hamano et al 2013). However, even early Earth may also have lost much of its volatiles in the first 100 Myr (Finlay et al 2016; Pujol et al 2013), with later water delivery occurring during the late veneer (Frank et al 2012).

### 3.1 Evidence for lost water

A tantalizing clue of the past nature of Venus lies in the deuterium-to-hydrogen ratio (D/H ratio) in its atmosphere. The D/H ratio can be indicative of fractionation by loss processes: lighterweight H is lost to space more easily than the heavier D. The D/H ratios of other solar system bodies are often quoted in comparison to Earth's Standard Mean Ocean Water (SMOW). Interestingly, the D/H ratio of SMOW is close to that of chondritic water (Alexander et al 2012), which may suggest the origin of some of Earth's water. The Venusian D/H ratio is estimated to be about 0.016 +/- 0.002 (Donahue et al 1982), or 157 ± 30 times terrestrial SMOW (Donahue et al 1982; 1997), although these measurements are somewhat complicated by the clogged instrument inlet on Pioneer Venus that obtained them. Thus, better measurements of the bulk D/H ratio of the Venus atmosphere will aid future studies of its potential early water inventory and loss processes. Measurements from the nightside infrared spectral windows suggest D/H ~ 127 x Earth's in the Venus near-surface environment (Bezard et al 2011), while VEx measurements above the clouds suggest that D/H may be 3x larger than Pioneer Venus suggested (Bertaux et al 2007).  Estimates from D/H suggest early Venus may have had 4-525 m of liquid water if spread evenly over surface (e.g. Donahue and Russell 1997), although other estimates suggest Venus formed with less water (Raymond et al 2006). An alternative explanation for the elevated D/H ratio in the Venus atmosphere is that it may reflect steady-state evolution where fractionating water loss is balanced by cometary or volcanic input fluxes (Grinspoon 1993; Donahue & Hodges 1992). Currently, the amount of $H_2O$ on Venus is $6 \times 10^{15}$ kg, which is much smaller than Earth's $1.4 \times 10^{21}$ kg (Lecuyer et al 2000). Below the clouds, the atmosphere contains only about 30 ppmv of water vapor (Chamberlain et al 2013; Arney et al 2014), and higher layers are even more dry (Figure 1).  Despite the desiccated conditions of Venus today, the tantalizing D/H ratio has been much-studied in the context of possible early habitable conditions.

Even at the start of its history, Venus would have been interior to the edge of the "conservative" habitable zone in the solar system (Kopparapu et al. 2013), so the potential existence of an early liquid water ocean is challenging to reconcile in light of this. However, recent 3-D modeling efforts have elucidated a path to a habitable, watery young Venus consistent with solar evolution (Way et al. 2016). Venus has a slow rotation period of about 116 Earth days (its sidereal day is 243 Earth days). On slowly rotating planets, atmospheric circulation patterns can generate thick substellar cloud decks caused by strong rising motion on the dayside. The thick water vapor clouds generated by this process produce substantial surface cooling by increasing the planet's dayside albedo (Yang et al 2016; Kopparapu et al 2017). It has been suggested that atmospheric tides acting on the thick atmosphere that the planet currently possesses lead to its present slow rotation rate (Dobrovolskis & Ingersoll, 1980), and so its current slow rotation is unlikely to be primordial. However, more recent work suggests that even a 1 bar atmosphere can create this tidally-induced rotation slowing (Leconte et al 2015). Way et al. (2016) shows that for a modern Earth-like atmosphere, rotation rates slower than 16x modern Earth's rotation rate can produce habitable conditions for early Venus, and at its present day rotation rate, habitable surface conditions are possible for up to at least 0.715 billion years ago. This means that Venus could have been habitable up to the time of its purported global resurfacing event. These simulations even show that snowfall could have occurred on the Venus nightside billions of years ago. The importance of high albedo water clouds for maintaining habitable conditions on Venus was also showed in a previous study by Grinspoon and Bullock (2007), who found that if the atmosphere of young Venus was cloud-free, it would have only been able to stave off catastrophic loss of habitability for roughly 500 million years after its formation. With 50% cloud coverage, climate stability can be maintained for up to 2 billion years after formation.

## 3.2 Loss of habitability

Even if Venus ever was habitable through mechanisms aided by its slow rotation rate, thick clouds, or other processes, it could not escape its inevitable fate. All main sequence stars steadily brighten as they age, pushing the boundaries of their habitable zones ever outward. As the sun slowly yet inexorably continued its steady march towards increasing luminosity, habitable conditions eventually could no longer be supported.

How did Venus go from a potentially clement world to our scalding planetary neighbor? As temperatures increased on Venus, driven by the brightening solar luminosity, increasing amounts of water would have evaporated into the atmosphere. Water vapor is a greenhouse gas, and so this would have created a positive feedback loop: higher temperatures caused the evaporation of more water, which in turn drove temperatures even higher, which in turn caused more water to evaporate, etc. As increasing quantities of greenhouse gases like water vapor are added to a planetary atmosphere, the outgoing thermal radiation emitted to space (also known as "outgoing longwave radiation" or OLR) will decrease. Thus, to balance the radiation budget of the atmosphere, the surface temperature must increase. There is a maximum limiting amount of OLR that a planet with a moist atmosphere can radiate to space known as the Kombayashi-Ingersoll limit. For an Earth-like atmosphere, the Kombayashi-ingersoll limit is about 300 W/m$^2$. Increasing quantities of greenhouse gases such as water vapor in a planet's atmosphere will trap increasing quantities of heat, but when the Kombayashi-Ingersoll limit is reached, the atmosphere is saturated with water and atmospheric spectral windows through which energy can be radiated to space become opaque, a maximally efficient greenhouse. This is known as a "runaway greenhouse" (Ingersoll 1969). In principle, a planet that has undergone a runaway greenhouse can heat up to such extreme temperatures that it is able to radiate in the visible part of the spectrum, where radiation can again escape to space. Ultimately, through dissociation of water vapor at high altitudes into hydrogen and oxygen, the water in the planet's atmosphere that has been evaporated from the surface can be lost as H escapes to space and O is lost through either top-of-atmosphere processes and/or reactions with surface minerals described below.

In practice, before a planet enters the runaway greenhouse regime, it can enter a state called the "moist greenhouse," at lower temperatures (Kasting 1988). On Earth, most of the water vapor in our atmosphere is "cold trapped" in the lowest atmospheric layer, the troposphere. The top of the troposphere is called the "tropopause," and here, water vapor condenses into clouds and rains back out to the surface. On Earth, this occurs at roughly P = 0.1 bar (e.g. Robinson & Catling 2014), or 9-20 km in altitude. Layers of the atmosphere above the tropopause are comparatively dry. However, as planetary surface temperatures increase and the atmosphere grows more and more moist with evaporated water vapor, the cold trap will lift to higher altitudes. As this occurs, the stratosphere of the planet can become "moist" with orders of magnitude more water vapor than it was possible for these altitudes to support when the cold trap was lower in the atmospheric column. Water at these high altitudes can be lost to space via dissociation by high energy radiation followed by the escape of hydrogen. The timescale for water loss approaches the age of Earth when the stratospheric water vapor mixing ratio is $3\times10^{-3}$, which occurs at a surface temperature of 340 K (Kopparapu et al. 2013). D/H fractionation only occurs efficiently for atoms transported to the exobase of the atmosphere; otherwise, relatively non-fractionating hydrodynamic escape processes will operate. Over time, as $CO_2$ liberated from rocks and volcanic outgassing entered the atmosphere, this more massive atmosphere could transport deuterium and hydrogen to higher altitudes where fractionating thermal and nonthermal escape processes could occur. In this scenario, this fractionating escape occurred only during the loss of the final 4 bar of $H_2O$ (Kasting 1988), so the D/H ratio remnant that is detectable today may not reflect the total initial water inventory. Kasting (1988) concluded that Venus could have maintained its oceans for 600 million years after its formation before losing them to the moist greenhouse, but this study excludes the effect of clouds that lengthen the interval of possible habitability (Section 3.1).

While free hydrogen can be easily lost to space, the fate of the liberated oxygen atoms from water photolysis is more ambiguous. The current Venus atmosphere does not host a significant quantity of free oxygen. The oxygen produced by water photolysis may have been lost to reactions with crustal and mantle materials. Chassefiere (1996) calculated that Venus' surface and mantle could have absorbed $O_2$ for initial water inventories up to 0.45x one Earth ocean. Additional ocean water beyond this could produce $O_2$ left behind in the atmosphere, resulting in an $O_2$-rich atmosphere (Section 4.3). Surface and mantle sequestration processes of $O_2$ are still not well understood, however (Rosenqvist & Chassefiere 1995). Unfortunately, because the entire surface of Venus is less than a billion years old, geological evidence of oxidation processes has likely been lost. At present, the modern Venus crust does not appear to be as oxidized as Mars (Pieters et al 1986).

As an alternative or in addition to surface and mantle sequestration of $O_2$, some fraction of the oxygen produced by water photolysis may have been lost to space. If Venus previously had a magnetic field, that may have played a role in slowing loss processes of volatiles from its atmosphere. However, it does not currently have a magnetic field, and gaseous species are still escaping. The Analyzer for Space Plasmas and Energetic Atoms (ASPERA-4) instrument aboard VEx has showed that dominant ions escaping from the modern atmosphere include H+, He+, and O+. Stoichiometrically, water is still being lost from Venus: the H+/O+ escape rate ratio measured by ASPERA-4 is 2:1 (Barabash et al 2007). More recently, analysis of VEx measurements have shown that Venus has an ambipolar electric field that can efficiently strip even heavy ions like oxygen to space (Collinson et al 2016). Surprisingly, the electric field surrounding Venus is five times stronger than the field in Earth's ionosphere. It is unclear why the Venus field is unexpectedly strong, but it may be because Venus is closer to the sun than Earth and thus is buffeted by larger amounts of ionizing radiation. This "electric wind" appears to be persistent, stable, and global in extent, and it can carry heavy O+ ions away to space even without solar wind stripping (Collinson et al 2016). Therefore, regardless of Venus' initial water inventory, its atmospheric loss processes for oxygen appear to be more efficient than they are for Earth.

On Earth, $CO_2$ outgassed into the atmosphere can, over long periods of time, be recycled back into the planet's mantle through reactions with silicates in Earth's crust that eventually are cycled back into the mantle via plate tectonic cycling and subsequently outgassed back into the atmosphere, completing the cycle. Indeed, this recycling mechanism has been used to explain Earth's relatively stable climate over geological timescales (e.g. Walker et al. 1981) because the rate of $CO_2$ reactions with silicate rocks is temperature-dependent, and so it produced a stabilizing negative feedback process. On Venus, the lack of plate tectonics means that $CO_2$ outgassed into the atmosphere has no way out and simply accumulates over time. The fact that Venus' atmosphere appears to contain a comparable amount of $CO_2$ to Earth's total $CO_2$ budget locked in carbonate rocks (Sleep & Zahnle 2001) suggests that Venus has outgassed all or most of its mantle $CO_2$ budget. Thus, the thick Venusian $CO_2$ atmosphere tells us of a long history of outgassing.

### 3.3 Noble Gases
Noble gases can record the early history of terrestrial planet evolution, so they are useful for constraining early states of planetary environments since they do not react with other gases or surface materials. Therefore, by analyzing noble gases on Venus, we could learn more about its early environment. Such gases include He, Xe, Kr, and Ar. Additionally, information about ancient geological processes may be discovered by measuring radioactive isotopes of noble gases (Zahnle 1993).

Zenon (Xe) isotopes observed on Mars and Earth exhibit strong mass fractionation, which may be evidence of atmospheric erosion processes that occurred during these planets' formation and early history (Pepin 1991; Zahnle 1993). Zenon has not been measured on Venus, but it could shed light on early atmospheric processes. Radiogenic isotopes of Xe on Earth and Mars suggest major catastrophic events (e.g. volatile-delivering comet impacts) on both planets occurring late in their formation (Pepin 2000). By understanding the Xe isotopes of Venus, we will better understand its early history and what source(s) may have delivered volatiles to it. If Venus has a similar Xe fractionation to Earth and Mars, this would suggest that all three bodies were delivered volatiles by the same source.

Measurements of Xe on Venus could also help to test a hypothesis that the source that endowed Earth with Xe was depleted in heavy isotopes. This is known as the "U-Xe" hypothesis (Pepin 1991). U-Xe would have a ratio of $^{136}$Xe/$^{130}$Xe that is 8% less than meteorites and the solar wind, and this has been invoked as a mechanism to explain Earth's Xe isotope ratios. U-Xe might be discovered on Venus, implying that it accreted Xe from the same source as Earth.

Pioneer Venus measurements suggest that Venus has an unusually high amount of Ne, $^{36}$Ar and $^{38}$Ar (Pepin 1991). These unusual abundances may suggest a large impact event early in the planet's history. For example, a large planetesimal endowed with noble gases from the solar wind could have impacted Venus and affected its abundances of these noble gases (McElroy and Prather 1981). As an alternative, the Ar and Ne observed on Venus might also have been delivered by a large comet (Owen et al 1992).

Radiogenic $^{40}$Ar could show evidence of degassing and surface processes. Venus has 25% less $^{40}$Ar than Earth. Because $^{40}$Ar is created by the decay of $^{40}$K, this could imply Venus has less K than Earth, or that it has for some reason outgassed less of this gas. Alternatively, Venus' inventory of $^{40}$Ar may be indicative of hydrological crustal weathering processes and may suggest that Venus experienced only one fourth of the amount of hydrological weathering experienced by Earth over the solar system's history (Watson et al 2007).

### 3.4 The future of Earth
Venus is a magic mirror that reflects the future of Earth back to forward-thinking observers. All main sequence stars brighten slowly over their lifetimes, and the inner edge of the solar system's habitable zone will someday sweep past Earth. The solar luminosity as a function of time can be estimated by (Gough 1981):

$$L(t) = \left[1 + \frac{2}{5}\left(1 - \frac{t}{t_0}\right)\right]^{-1} L_0$$

Extrapolating into the future, Wolf and Toon (2015) find that stable climates on Earth can be maintained for solar insolations < 21% greater than today and surface temperatures < 368.2 K. Climate modeling by Wolf and Toon (2015) suggests these conditions will be met about 2 billion years into the future. However, loss of habitability could occur before that. Our planet's water could be lost in as little as ~130 million years in the future before a thermal runaway can occur. Along with water loss and surface heating, a shutdown of plate tectonics should occur in future. Subsequent volcanic outgassing of $CO_2$ and $SO_2$, coupled to photochemical processes that generate $H_2SO_4$ clouds, could lead to a state closely resembling Venus on Earth.

**4 Venus: The Exoplanet Laboratory Next Door**
A major focus of astrobiology is the search for habitable conditions and life beyond Earth. We have scrutinized Earth intensely to understand how we might search for habitable conditions elsewhere. In many ways, Venus is the most Earth-like planet in the solar system with its similar mass, radius, and bulk density. Indeed it is likely that Venus and Earth had very similar starting conditions in terms of their relative compositions of both volatiles and refractory compounds. Yet Earth has been habitable since at least the start of the Archean eon and possibly during the Hadean eon (Bell et al 2015), while Venus' timescale for habitability is much more uncertain. At some point, the evolution of these two planets diverged dramatically, and Venus is now one of the most uninhabitable planets we might imagine.

In the same way that we can study Earth's history to understand of how biospheres may evolve over time and co-evolve with their environments, Venus can teach us about an equally fundamental process: how planets reach the end-state of planetary habitability. While Earth shows how self-regulating negative feedbacks (e.g. the carbonate silicate cycle) work to maintain and preserve habitability on geological timescales, Venus offers us an example of a world whose self-regulation mechanisms failed catastrophically. Would we ever have imagined that a world like Venus could exist among the exoplanets if we did not have an example of one nearby? Without Venus in our solar system, we would be dramatically more inclined to consider the size and mass of the Earth as a fundamental aspect of driving habitable conditions in terms of both the geological and atmospheric evolution. Yet Venus shows us that knowing the size and mass of an exoplanet is not only insufficient for quantifying habitability, but it demonstrates that a similar planetary structure to the Earth can produce an environment that may be considered at the opposite end of the habitability scale. Therefore, in our searches for worlds circling distant stars, an incomplete understanding of the evolution of Venus and the processes that govern its atmosphere and interior today will hinder our ability to interpret observations of hot terrestrial exoplanets, and will also hinder our ability to understand processes that govern planetary habitability more generally.

Many basic parameters of Venus are still unknown. These include the relative size of the core to its mantle, the rate of current volcanic activity, its moment of inertia, and the identity of its unknown UV absorber. Yet Venus accounts for 40% of the mass of the rocky solar system planets and is the closest planet in the universe to Earth. When we work to understand the basic properties of exoplanets, we must have a robust understanding of our local planetary neighborhood to guide us. For example, comparisons between Earth and solar compositions are being used to predict the interiors of exoplanets based on the stellar abundances of the host star (Hinkel & Unterborn 2018). Such inferences of planetary abundances will benefit enormously from a more detailed understanding of those abundances present within the terrestrial planets of our solar system. Moreover, as we describe here, Venus-like exoplanets may be one of the most common types of terrestrial worlds, and our current and near-future detection and characterization techniques are biased towards detecting exo-Venus analogs over exo-Earth analogs.

Because the primary exoplanet discovery techniques, such as the radial velocity and transit methods, are biased to detecting exoplanets hotter than Earth, much of the modeling work that has yet been undertaken of exoplanet environments has focused on understanding hot worlds (e.g. Schaefer and Fegley 2009, 2011; Leger et al 2009; Miguel et al 2011; Treiman and Bullock 2012). Very hot planets (i.e. $T_{surf}$ > 1500 K) are likely to be in a regime even beyond Venus, with atmospheres composed of exotic compounds including SiO, and clouds made of K and Na compounds (Schaefer and Fegley 2009). Hot planets in a less extreme temperature regime may be more like Venus, with $CO_2$ atmospheres and other compounds that may include $N_2$, $H_2O$, $SO_2$, HCl, HF, OCS, and CO (Schaefer and Fegley 2011; Treiman and Bullock 2012), all of which occur in the atmosphere of Venus.

### 4.1 Exo-Venus Analogs

The development of climate models over recent years have allowed us to estimate the inner and outer edges of the Habitable Zone (HZ) for different mass main sequence stars (Kasting et al 1993; Kopparapu et al 2013, 2014). Importantly, these HZ calculations allow us to estimate the fraction of stars with rocky planets within the habitable zone (referred to as $\eta_{Earth}$). Recent calculations of $\eta_{Earth}$ allow us to predict how many potentially Earth-like planets we may detect with future observatories. Much of these recent calculations of $\eta_{Earth}$ use Kepler statistics to provide a sufficiently large sample to perform a meaningful statistical analysis (Dressing & Charbonneau 2013, 2015; Kopparapu 2013; Petigura et al. 2013).

As previously mentioned, the transit method has a dramatic bias towards the detection of planets which are closer to the host star than farther away (Kane & von Braun 2008). Additionally, a shorter orbital period will result in an increased signal-to-noise (S/N) of the transit signature due to the increased number of transits observed within a given timeframe. The consequence of these factors are that *Kepler* has preferentially detected planets interior to the HZ, which are therefore more likely to be potential Venus analogs than Earth analogs. Since the divergence of the Earth/Venusian atmospheric evolutions is a critical component for understanding Earth's habitability, the frequency of Venus analogs ($\eta_{Venus}$) is also important to quantify. Observing the boundary between Venus-like worlds and Earth-like worlds will also allow us to place empirical constraints on the interior edge of the habitable zone. It may even allow us to quantify the likelihood of processes that could help to maintain habitability inside the inner edge of the HZ (e.g. thick substellar cloud decks, as has been invoked to explain

Venus' apparent ability to support liquid water for some unknown interval of time after its formation, Section 3.1).

Kane et al. (2014) defined the "Venus Zone" (VZ) as a target selection tool to identify terrestrial planets whose atmospheres could be pushed into a runaway or moist greenhouse, producing environmental conditions similar to those on Venus. Figure 3 shows the VZ (red) and HZ (blue) for stars of different temperatures. The outer boundary of the VZ is the "Runaway Greenhouse" line, which is calculated using climate models of Earth's atmosphere. The inner boundary (red dashed line) is estimated based on where the stellar radiation from the star would cause complete atmospheric erosion. The images of Venus shown in this region represent planet candidates detected by Kepler that may be Venus-analogs. Kane et al. (2014) calculated an occurrence rate of VZ terrestrial planets as 32% for low-mass stars and 45% for Sun-like stars. Note however that, like the HZ, the boundaries of the VZ should be considered a testable hypothesis since runaway greenhouse could occur beyond the calculated boundary (Foley 2015; Luger and Barnes 2015, Section 4.2).

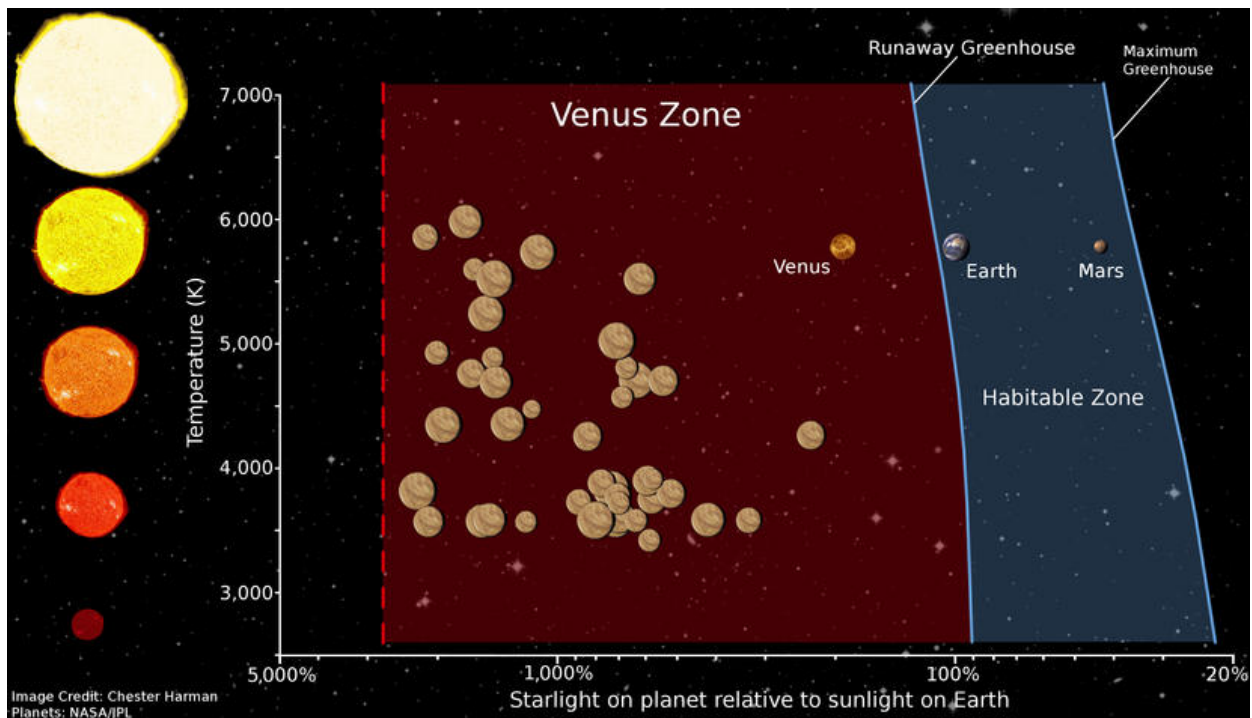

**Figure 3:** The boundaries of the classical "Habitable Zone" and the "Venus one", represented as a function of stellar temperature and insolation flux relative to the Earth. Shown on the plot are the location of the solar system planets (Venus, Earth, Mars) and the potential Venus analogs identified by Kane et al. (2014).

The prevalence of Venus analogs will become increasingly relevant in the era of forthcoming exoplanet missions. The Transiting Exoplanet Survey Satellite (TESS) is expected to detect hundreds of terrestrial planets orbiting bright host stars (Sullivan et al. 2015), many of which will lie within their stars' VZ. The PLAnetary Transits and Oscillations of stars (PLATO) mission will add further to the inventory of candidate Venus analogs orbiting bright stars and potentially extend to longer orbital period sensitivity than TESS (Rauer et al. 2016). These new discoveries will provide key opportunities for transmission spectroscopy follow-up observations using the James Webb Space Telescope (JWST), among other facilities, such as the Atmospheric Remote-sensing Exoplanet Large-survey (ARIEL) mission (Zingales et al. 2018). Observations capable of identifying key atmospheric abundances for terrestrial planets will face the challenge of distinguishing between possible Venus and Earth-like surface conditions and also understanding the data to correctly understand the planetary environment. Discerning the actual occurrence of Venus analogs will help us to decode why the atmosphere of Venus so radically diverged from its sister planet, Earth, and will help to constrain how frequently these processes occur.

We may already have detected several of Venus' sisters in the exoplanet population. These worlds include planets such as TRAPPIST-1 b and c (Gillon et al 2016, 2017), which receive 4.3 and 2.3 times Earth's insolation; Kepler-1649b (Angelo et al. 2017), which receives 2.3 times Earth's insolation, GJ 1132b (Berta-Thompson et al 2015), which receives 19 times Earth's insolation; and Ross 128b (Bonfils et al. 2017), which receives 1.38 times Earth's insolation. True Venus-analogs would display relatively featureless spectra, both in transit transmission and direct imaging. So, while these worlds may be cosmically ubiquitous, they may be challenging to observe and interpret. This underscores the importance of allowing the Venus in our solar system to guide our understanding of exo-Venus worlds. Figure 4 shows the spectrum of Venus for reflected and transmitted light.

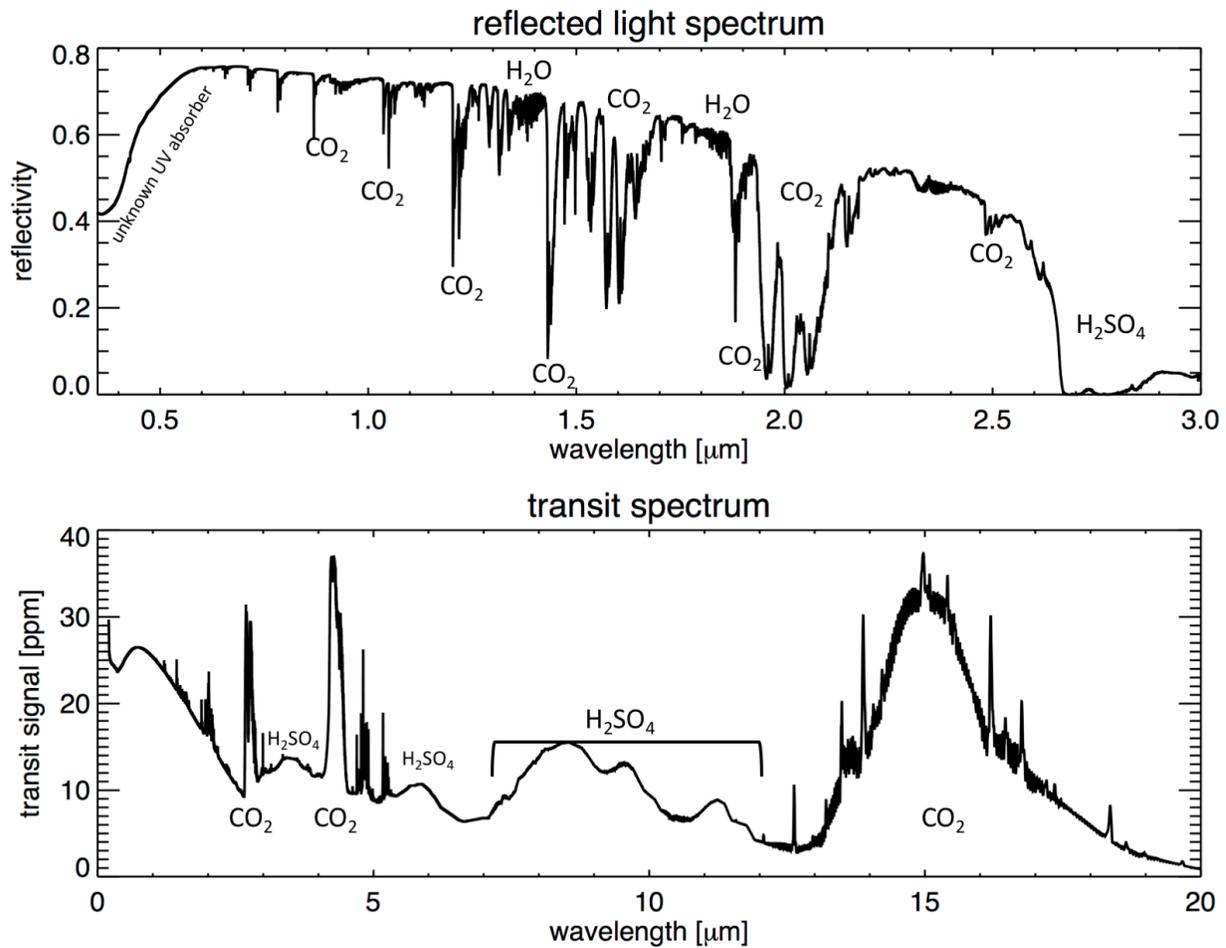

**Figure 4.** The reflected light spectrum (top panel) and a transit transmission spectrum of a Venus-analog orbiting an arbitrary M dwarf star (bottom panel).

In reflected light (top panel of Figure 4), absorption features can be discerned from $CO_2$, and for wavelengths < 0.5 μm, the unknown UV absorber is prominent, darkening the spectrum. At wavelengths > 2.5 μm, $H_2SO_4$ becomes significantly absorptive (Figure 2), causing the spectral falloff in the reflected light spectrum. Although water vapor is a trace species in the Venus atmosphere, it is present throughout the atmospheric column, and water vapor features can be seen in the reflected light spectrum. In transit transmission, the cloud deck dramatically truncates the strength of gaseous absorption features, but it is still possible to observe $CO_2$ features. $H_2SO_4$ itself also produces absorption features that could be observed in transmission observations, notably near 10 μm. These features might be detectable with JWST, though JWST may encounter a noise floor of 20 ppm in the near-infrared and 50 ppm for mid-infrared instruments (Greene et al. 2016). The actual impact of this noise floor will not be known until after telescope commissioning and early observations are completed. Fortunately, planets orbiting smaller stars like TRAPPIST-1 will produce larger transit depths, possibly with some features exceeding 100 ppm (Morley et al. 2017) for clear atmospheres. Clouds will decrease the strength of absorption features, but strong features (e.g. the $CO_2$ bands near 4 and 15 μm) might still be detectable on exo-Venuses. Because observations with JWST will be costly in terms of observing time, careful target selection is absolutely critical (Morgan et al. 2018).

Venus-like planets may also be observable through thermal phase curves (Morley et al 2017; Meadows et al 2018). Such observations may be possible through photometry using the JWST Mid Infrared Instrument (MIRI). The planet-star contrast ratio increases at longer mid-infrared wavelengths. Observations of the 15 μm $CO_2$ band at thermal wavelengths may provide information that could help to discriminate between different types of planetary environments (and also act as a discriminant between planets with atmospheres and those without; Meadows et al 2018). If the planet's emitted flux at 15 μm is lower than the expected thermal emission continuum level based on other wavelengths, this is evidence for $CO_2$.

The clouds dramatically impact the spectrum of Venus -- including producing spectral features of their own (Figure 4) --  and it will be important to better understand and predict the potential for clouds to form in exo-Venus atmospheres to anticipate their impact on observations. Because the $H_2SO_4$ in Venus' atmosphere that is the source of the clouds is generated by photochemistry, the UV spectrum of the star will impact how much $H_2SO_4$ is generated on exoplanets. Possible, planets around stars with higher UV outputs may develop thicker cloud decks (Schafer and Fegley 2011), which will have important implications for the detectability of spectral features on exo-Venus analogs since thicker clouds, and higher cloud decks, lead to diminished gaseous spectral features.

**4.2 Exo-Venus Analogs in the Habitable Zone**
In addition to exoplanets in the Venus Zone, some exoplanets in their stars' HZ may in fact be more Venus-like than Earth-like for worlds orbiting M dwarfs. Planets in the HZs of M dwarfs may experience significantly different evolutionary histories compared to planets orbiting solar-type stars. M dwarfs, especially younger stars, are highly active, producing frequent flares (West et al 2015; MacGregor et al 2018) and high amounts of x-ray radiation that can erode atmospheres significantly (e.g. Owen & Mohanty 2016; Airapetian et al. 2017; Garcia-Sage et al. 2017; Dong et al 2017). M dwarfs are also inherently dim compared to solar-type stars, and planets must huddle close to them for warmth, making them vulnerable to this extreme activity and strong tidal forces (Barnes et al., 2013).

Even before planets orbiting M dwarfs can settle into their long lifetimes in the habitable zone, they must contend with their stars' extended super-luminous pre-main sequence phase caused by longer Kelvin-Helmholtz contraction timescales for lower mass stars (e.g., Baraffe et al., 1998; Reid and Hawley, 2013; Dotter et al., 2008). A one solar mass star contracts to the main sequence in < 50 million years (Baraffe et al 1998), while M dwarfs can spend hundreds of millions of years to up to a billion years contracting. During this phase, stars can be up to two orders of magnitude more luminous than their eventual main sequence brightness, so planets that orbit within the ultimate habitable zone of M dwarfs may be well interior to it for an extended period of time. Studies suggest that rocky planets form in 10 - 100 million years (Chambers, 2004; Raymond et al., 2007, 2013; Kleine et al.,2009), so planets in M dwarf habitable zones are subjected to extreme luminosities early on for an extended period of time. This lengthy extreme luminosity can cause these planets to enter runaway greenhouse states before their stars stop contracting and the HZ boundaries settle to their main sequence phase positions. Therefore, planets orbiting in the habitable zones of M dwarfs may actually be desiccated exo-Venus worlds in disguise despite their orbital parameters suggesting potential habitability (Luger & Barnes 2015; Ramirez & Kaltenegger 2014; Meadows et al., 2018). Models show that planets enduring the extended super-luminous pre-main sequence phase of an M dwarf could lose several Earth ocean equivalents of water through evaporation and hydrodynamic escape. Therefore, the "Venus zone" may extend into the HZ for some M dwarfs. Planets in the outer parts of early M dwarf HZs (i.e. M1 - M3V) are less likely to become exo-Venus planets than worlds orbiting closer to the inner edge of the HZ and/or orbiting later M dwarfs. Some of these potential exo-Venus habitable zone planets include TRAPPIST-1 d-g (M8V), Proxima Centauri b (M5.5V), and LHS 1140b (M4.5V). Others will surely be discovered by TESS.

It is possible that habitability could be restored to planets desiccated by the pre-main sequence phase if late volatile delivery occurs (Morbidelli et al. 2000). Alternatively, planets could migrate into the HZ from farther out in the system where solid ices could endow them with abundant volatiles, and complete desiccation might be averted. Indeed, the resonant orbits of the planets in the TRAPPIST-1 system suggests that radial migration within a gaseous disk occurred (Luger et al 2017). The migration timescale for this system was likely much shorter than the pre-main sequence phase of TRAPPIST-1 (up to 1 billion years; Bolmont et al 2016), so a "late" migration that might avoid being close to the star during the damaging pre main sequence is unlikely. However, estimates of the density of the TRAPPIST-1 planet suggest they are endowed with a significant volatile fraction, which may have allowed them to avoid total desiccation (Unterborn et al. 2018). TRAPPIST-1 f and g may contain > 50% water by weight, while TRAPPIST-1 b an c are drier, but probably still more volatile rich than Earth ($\leq$ 15% water by weight). In any case, the extended pre-main sequence phase for M dwarfs suggests that Venus-like states are possible for habitable zone planets, depending on initial planetary volatile inventory (provided these planets can retain atmospheres despite the activity of their stars). The types of scenarios that can "save" these planets from desiccation and how frequently they occur are just beginning to be explored, and observations of HZ exoplanets orbiting M dwarfs should provide valuable insights.

As discussed in section 3.1, Venus' slow rotation rate may have enabled habitability for an extended period of time by forming a thick layer of water clouds on the dayside driven by vigorous convection at the substellar point. Beyond the solar system, slow rotation may be likely for synchronously-rotating exoplanets within or inside the HZs of M dwarf stars. Similar thick substellar cloud decks may affect the potential habitability of these worlds, cooling their surface temperatures and possibly allowing for habitable conditions interior to the inner edge of the traditional HZ (Kopparapu et al 2017; Fujii et al 2017). Such planets may even be in a "habitable moist greenhouse" states with moist stratospheres gradually losing water to space, but with habitable surface temperatures. If water loss is slow, habitable conditions may exist in these planets for a significant period of time (e.g. hundreds of millions to several billion years;

Kopparapu et al 2017). By studying the history of Venus, we can gain a better understanding of how these processes may operate elsewhere. Likewise, by observing habitable slowly-rotating exoplanets, this can help us to understand the potential for this "habitable moist greenhouse" process to have occurred on Venus in the past. In light of the difficulty of obtaining data from the early history of Venus, these observations of analog exoplanets are particularly valuable for illuminating the history of Venus (Section 2.3).

**4.3 Exo-Venus Analogs and False Positive Oxygen**
One consequence of the extreme water loss that planets endure during a runaway or moist greenhouse could be the accumulation of significant quantities of $O_2$ in planets' atmospheres (Luger and Barnes 2015; Tian 2015). Depending on the initial water inventory, hundreds or thousands of bars of $O_2$ could be generated. Photolysis and hydrogen loss from one Earth ocean would produce about 240 bars of $O_2$ (Kasting 1997). The amount of $O_2$ that actually remains in the atmosphere would be a function of a number of processes including stellar activity, the original quantity of water, the mass of the planet, oxygen sinks, and the planet-star distance.

Oxygen has been heavily studied in the context of atmospheric biosignatures (e.g. Hitchcock and Lovelock 1967; Meadows et al 2017). Therefore, any "false positive" processes that generate $O_2$ abiotically are important to consider when planning for future observations of potentially habitable exoplanets designed to search for signs of life. In light of this, several previous studies have discussed $O_2$ biosignature false positives (Hu et al., 2012; Domagal-Goldman et al., 2014; Tian et al., 2014; Gao et al., 2015; Harman et al.,2015; Schwieterman et al 2016; Meadows et al 2017; Wordsworth & Pierrehumbert 2014; Schaefer et al 2016).

Atmospheres with abundant oxygen produce spectral features from the UV to the mid-infrared from $O_2$ itself, from its photochemical byproduct ozone ($O_3$), and from collisional pairs of $O_2$ molecules called "$O_2$ dimers." Dimer features are collisionally-induced and appear mostly strongly in atmospheres with significant quantities of $O_2$. $O_2$ itself produces strong spectral features in the visible and NIR at 0.628 μm (the "B Band") and 0.762 μm (the "A band"), and the $a\,^1\Delta_g$ band near 1.27 μm (the designations "B band" and "A band" were first given by Joseph von Fraunhofer in his pioneering observations of the spectrum of sunlight passing through Earth's atmosphere). Of these features, the "A band" is the most prominent (Rothman et al 2013). Ozone produces strong absorption in the UV (0.2-0.3 μm, the "Hartley band"), the visible (0.5-0.7 μm, the "Chappuis band"), and in the MIR (9.6 μm). Of these, the UV "Hartley band" is the strongest and detectable for the lowest quantities of $O_2$ and $O_3$. $O_2$ dimer features are produced between 0.3 - 0.7 μm, at 1.06 μm, and at 1.269 μm (coinciding with the $a\,^1\Delta_g$ feature). The type of massive $O_2$ accumulation that could occur during the loss of oceans' worth of water should reveal itself as a biosignature false positive through the appearance of $O_2$ dimer features because these collisionally-induced absorption features only become prominent when a substantial amount of $O_2$ is present (i.e. bars). In transit transmission observations, for planets with up to 10 bars of pressure but with small $O_2$ atmospheric fractions (< 20%), $O_2$ dimer features are challenging to discern (Misra et al 2014). However, for a ~1 bar atmosphere dominated by $O_2$, the NIR dimer features are stronger (Schwieterman et al 2016). For the extreme case of a 100 bar $O_2$ atmosphere (which could result from the loss of 0.4 x Earth's ocean volume), the NIR $O_2$ dimer features become stronger than the $O_2$ features themselves, a clear spectral indicator that implies the abiotic nature of the oxygen. In direct imaging, the dimer features at both visible and NIR wavelengths can be detected. In transit transmission, Rayleigh scattering tends to mask visible wavelength dimer features due to the longer path lengths through the atmosphere.

As one case study of a potential extreme kind of exo-Venus, Schaefer et al (2016) modeled ocean loss for GJ 1132b in the presence of a magma ocean. This planet is well inside its star's habitable zone and receives a very high amount of stellar flux even compared to Venus (roughly 19x Earth's insolation, or 9.5x Venus' insolation). Although this planet's temperature regime is even higher than Venus, it is still instructive to consider the possibility of abiotic $O_2$ on it as an extreme scenario. Schaefer et al suggest that the most likely outcome for water loss on GJ 1132b is a thin $O_2$ atmosphere, although large initial water abundances could produce atmospheres with 1000s of bars of $O_2$.

Besides the super-luminous pre-main sequence phase driving massive water loss and the potential subsequent generation of $O_2$, other oxygen false positive scenarios have been proposed. These include the following scenarios, all of which can be better understood when considered in the context of Venus atmospheric processes as we will discuss below.

Low non-condensable gas inventories (e.g. low $pN_2$) can generate false positive $O_2$. This mechanism involves low pressure atmospheres that permit enhanced water vapor abundance at high altitudes because at low pressure, the atmospheric "cold trap" that would otherwise prevent water from reaching upper layers is less effective (Wordsworth and Pierrehumbert 2014). As we have discussed in the context of Venus' water loss, water vapor in the stratosphere is vulnerable to photolysis followed by loss. If the $O_2$ generated by $H_2O$ photolysis is not itself lost, it could produce detectable false positive spectral signatures. For a planet with modern Earth's surface temperature, reducing the atmospheric partial pressure of $N_2$ by an order of magnitude could allow for a moist upper atmosphere, and photochemical production of abiotic $O_2$ could follow (Wordsworth and Pierrehumbert 2014). There is evidence that the pressure of Earth's atmosphere may have been as low as 0.23 +/- 0.23 bar at 2.7 billion years ago (Som et al 2016). Interestingly, Tomkins et al (2016) analyzed 2.7 billion year old meteorites from this same time period on Earth and suggested that they show evidence of abundant stratospheric oxygen.

Abiotic oxygen can also be produced by other photochemical processes. Carbon dioxide photolysis can produce oxygen via:

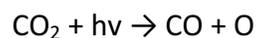
$$CO_2 + h\nu \rightarrow CO + O$$

This also occurs on Venus. Direct recombination of CO and O back into $CO_2$ is spin forbidden and proceeds very slowly. So, O produced by this mechanism can react to form $O_2$ and $O_3$ (Harman et al 2015; Domagal-Goldman et al 2014; Tian et al 2014; Hu et al 2012; Gao et al 2015). Therefore, atmospheres with abundant $CO_2$ and efficient $CO_2$ photolysis (which occurs via photons with wavelength < 175 nm) may show signs of false positive oxygen. However, catalysts present in an atmosphere may allow the recombination of CO and O back to $CO_2$. For instance, water vapor photolysis that generates OH and HOx chemistry can enable this through the reactions (Meadows 2017):

$$O + O_2 + M \rightarrow HO_2 + M$$
$$O + HO_2 \rightarrow OH + O_2$$
$$OH + CO \rightarrow CO_2 + H$$

The net result of this is $CO + O \rightarrow CO_2$. Gao et al (2015) pointed out that extremely desiccated atmospheres ( < 1 ppm $H_2O$, less than Venus) can enhance abiotic oxygen accumulation by inhibiting this catalytic recombination to $CO_2$. A lack of $H_2O$ features in combination with detectable oxygen and abundant $CO_2$ might suggest the production of $O_2$ from a highly desiccated planet (and also would indicate a planet that is uninhabited as water is considered fundamentally important for life).

Besides water vapor, oceans can also provide sinks to remove oxygen from the atmosphere: $O_2$ and CO in the ocean can reform $CO_2$ through aqueous chemistry (Harman et al 2015). However, stars with efficient $CO_2$ photolysis may still generate false positive $O_2$ even for planets with oceans if other $O_2$ sinks such as reduced volcanic gases (e.g. $CH_4$, $H_2S$) are minor components of the atmosphere. These types of false positive planets could be discriminated from planets with biological $O_2$ through the presence of strong $CO_2$ and CO features, and weak or nonexistent spectral features from reduced gases.

Venus has a vital role to play in guiding our understanding of the potential for the false positive $O_2$ scenarios discussed here to operate on exoplanets. Venus may have undergone massive water loss in its past, yet it does not exhibit an $O_2$-rich atmosphere today. The strong electric wind at Venus (Collinson et al 2016; Section 3.2) has significant implications not just for the potential for $O_2$-rich atmospheres to exist on exoplanets that have undergone massive water loss (e.g. through a runaway or moist greenhouse), but also for atmospheric retention more generally. Venus has an electric potential drop of +12 V in its ionosphere, and if exoplanets have a comparable or stronger field, they too could lose heavy ions like O+ to space regardless of whether they possess a magnetic field. The unexpected strength of the Venusian electric wind has been suggested to be due to Venus' closer proximity to the sun than Earth, causing it to receive more ionizing radiation. Planets orbiting closely to M dwarfs may have even stronger electric winds when considering their proximity to their stars and the high levels of stellar activity typical for M dwarfs. The potential for electric winds and other atmospheric loss processes on exoplanets must be investigated in detail to assess their impact on atmospheric retention, habitability, and biosignature false positive scenarios.

On Venus today, atmospheric $CO_2$ is photolytically dissociated (McElroy et al 1973). Photochemical production of abiotic $O_2$ and $O_3$ could therefore occur. Montmessin et al (2011) has reported small amounts of ozone ($1.5 \times 10^{-4}$ times Earth's ozone abundance) on the Venus nightside. Also on the nightside, recombination of excited $O_2$ produces airglow (Figure 5) that can be detected near 1.27 µm (Connes et al., 1979; Crisp et al 1996). This free oxygen is produced by $CO_2$ photolysis on the planet's dayside whose products circulate to the planet's nightside and recombine as excited $O_2$ (a$^1\Delta$g) in the mesosphere, which quickly relaxes to the ground state and emits the 1.27 µm photons detectable from space. The airglow produced by $O_2$ emission on Venus is highly variable, both spatially and temporally. Despite the presence of $O_2$ airglow from excited oxygen, ground-state $O_2$ has never been detected on Venus, with upper limits suggesting <2 ppm above 60 km (Trauger and Lunine, 1983; Mills, 1999). The implication is that the oxygen formed by $CO_2$ photolysis is rapidly removed from the atmosphere. Chlorine catalysts have been suggested as a mechanism to scrub $O_2$ from Venus' atmosphere (Yung and DeMore 1982); laboratory experiments (Pernice et al 2004) and measurements of Venusian $O_3$ abundance (Montmessin et al 2011) suggest this mechanism is viable, but intermediate species involved have not been detected in Venus' atmosphere. Alternatively, or in addition to catalytic chlorine chemistry, heterogeneous aerosol chemistry may help to scrub $O_2$ from the Venusian atmosphere (Mills and Allen 2007).

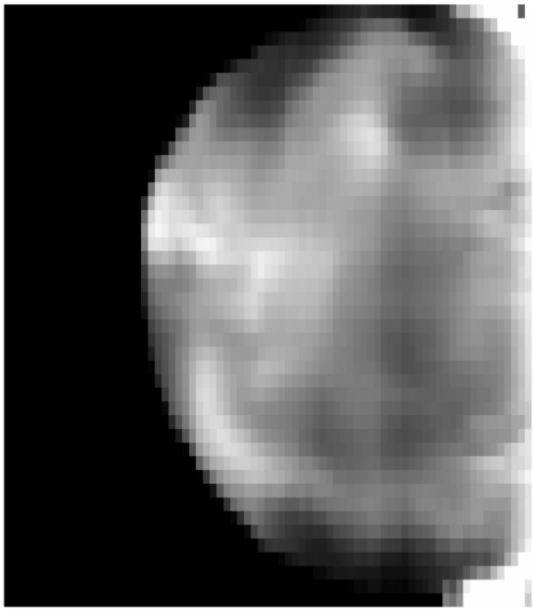 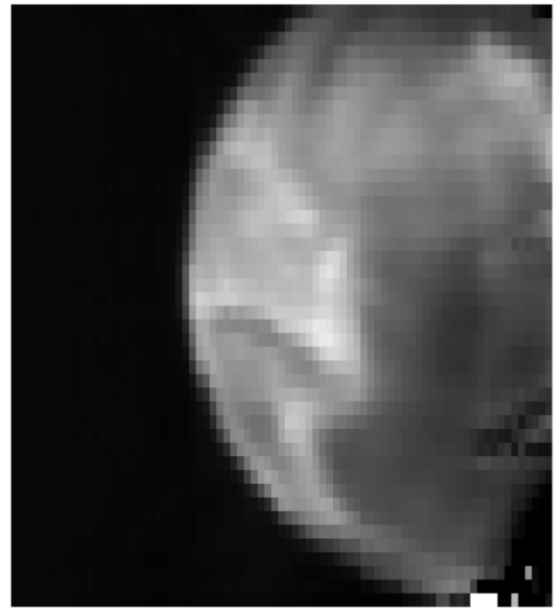

**Figure 5**. Variable $O_2$ airglow at 1.27 µm can be seen illuminating the Venus nightside in these observations from March 2, 2009 and March 3, 2009 from the Apache Point Observatory 3.5 m telescope TripleSpec instrument. The dayside crescent is to the right in these images.

It is noteworthy that the mechanisms that remove $O_2$ from the Venus atmosphere have not been included in studies examining the possibility of abiotic $O_2$ production on exoplanets as a biosignature false positive. This is in large part because these processes not still not well understood. Therefore, better understanding the processes involving $O_2$ in the Venusian atmosphere are vitally important for understanding the viability of many proposed oxygen false positive mechanisms for exoplanets.

## 5 Concluding Remarks

In 1918, Nobel prize winning physicist and chemist Svanthe Arrhenius penned *The Destinies of the Stars*, in which he supposes that Venus is a hot, steamy world of rainstorms and swamps. Waxing poetic on "radiant Venus" with its "brilliantly white lustre," he notes that this world has captivated our attention since ancient times. Owing to the high Venusian surface temperatures and abundant moisture, he reasons that "everything on Venus is dripping wet", and suggests that Earth's sister planet is home to "luxuriant vegetation." But just as modern Venus is simultaneously a world of astonishing beauty and incredible malevolence, it seems appropriate that Arrhenius quickly spoils his resplendent imagery as he goes on to suppose that the "dead, decaying bodies" of the short-lived "low forms of life" that he imagines make their home there fill the air with "stifling gases" as they disintegrate. And if they are unlucky enough to become entrapped in the rivers oozing with slime (what else?), they "speedily turn into lumps of coal." For Venus, what could be more appropriate than the grisly commingled with the sublime?

Venus teaches us that habitability is not a static state that planets remain in throughout their entire lives. Habitability can be lost, and the runaway greenhouse is the final resting place of once watery worlds. Understanding habitability as planetary process crucially depends on understanding what happened to the putative lost water of Venus. Although the fanciful humid swamps imagined by Arrhenius could not be farther from the desiccated reality of the current Venusian environment, perhaps they contain a grain of truth about past Venus.

Venus is extremely valuable in the context of exoplanets. We can make models that seek to understand exoplanet atmospheres and environments more robust by validating them across the diverse and complex atmospheres that exist in the solar system (e.g. Venus, Earth, Mars, etc). Given that the planetary atmosphere and interior models that we extrapolate and apply to exoplanets are based on in situ data acquired within our solar system, it is imperative that the planetary objects that are within reach are exploited to the fullest extent. As a world of extremes in temperature and pressure, Venus is particularly useful for model validation across a range of conditions. Because exoVenus planets may be cosmically ubiquitous, it is particularly important to better understand the world next door so that we may be able to better interpret future observations of analog worlds. Venus may even teach us something about detecting life on exoplanets by guiding our understanding of oxygen biosignature false positives. Likewise, just as Venus can teach us about exoplanets, studying hot exoplanets of varying ages and in varied astrophysical contexts may shed light on processes that have operated on Venus in the past. We may even observe "habitable moist greenhouse" planets interior to their stellar systems' habitable zones. The discovery of such worlds would prove that habitable past conditions for Venus are possible.

Near the end of his section on Venus, Arrhenius concludes with these thoughts: "The planets do tell us the conditions that existed on the Earth at the first dawn of life and we can also draw from them a prediction of the fate that once, after milliards of years perhaps, will befall the later descents of the present generation." These musings seem startlingly prescient when considering that exoplanets may someday allow us to witness planetary states lost to our solar system's past, and Venus may show us the possible future of Earth.